\documentclass[pra,twocolumn,amsfonts,amssymb,amsmath,floatfix,floats,a4paper]{revtex4-1}

\usepackage{graphicx}
\usepackage[dvipsnames]{xcolor}
\begin{document}

\title{Do neutrons disagree with photons  about where they have been inside an interferometer? }

\author{Lev Vaidman}
\affiliation{Raymond and Beverly Sackler School of Physics and Astronomy\\
 Tel-Aviv University, Tel-Aviv 69978, Israel}

\begin{abstract}
Recent experiments with identically tuned nested Mach-Zehnder interferometers which attempted to observe the location of particles inside these interferometers are analyzed. In spite of claims to the contrary, it is argued that all experiments support the same surprising picture according to which the location of the particles inside the interferometers is not described by continuous trajectories.
      \end{abstract}
\maketitle

\section{Trajectory of a quantum particle}

The trajectory of a quantum particle inside an interferometer is a controversial topic. It is easy to answer the question: `In which arms  the  wavefunction of the particle entering the interferometer appears?' However, even the meaning of the question: `Where was the pre and postselected particle inside the interferometer?' is not clear. Adding  strong measurements which would unambiguously tell us the particle path spoils the interference. So the results of such measurements do not necessarily help us to know where the particle was in the interferometer without intermediate strong measurement.

A seemingly reasonable approach was suggested by Wheeler \cite{Wheeler}, who proposed that when in the interferometer the original wave packet splits into several wave packets, only one of which reaches the detector, then the trajectory of this wave packet is the trajectory of the particle. He made this proposal in the context of  a delayed choice scenario which included a controversial claim about affecting the past, but the proposal was considered uncontroversial when there were no changes in the interferometer during the photon flight.

When a classical particle moves along  a particular trajectory it is expected that it leaves some (maybe very faint) trace in the environment, so a natural assumption is that this faint trace can be used as another definition of particle's  trajectory. A moving localized wavepacket of a quantum particle  is also expected to leave a faint trace: tiny local changes of the quantum states of the environment. These changes are not strong enough to leave robust records along the trajectory of the particle in the form of quantum states orthogonal to the states of an undisturbed environment. The faint traces are small amplitudes of orthogonal (to the undisturbed states) components which can be found only from experiments on an ensemble of particles with the same pre and postselection. Even these faint traces can be taken as definitions of the trajectory of the particle: {\it the particle was where it left a trace of the order a localized wave packet would leave}. In the case of the motion of a well localized packet this definition provides  a continuous line starting at the source and ending at the detector.

\section{Nested Mach-Zehnder interferometer}

In most cases the trace definition and Wheeler's proposal provide the same pictures of trajectories. However, it was found \cite{past} that for a particularly tuned nested Mach-Zehnder interferometer (MZI), the two proposals yield different pictures. According to Wheeler, the trajectory is a single continuous line. The faint trace will be on all parts of this line, but also in a separate loop disconnected from the line. The prediction of this closed loop, which does not start at the source and does not end at the detector, generated a hot discussion \cite{LiCom,RepLiCom,morepast,Bart,BartCom,Poto,PotoCom,Grif,GrifRep,Hash,HashCom,HashComRep,Dupr,DuprCom,DuprComRep,Eli,Disapp,Sokol2,ACWE} which only intensified \cite{Danan,Saldan,Jordan,JordanCom,Sali,SaliCom,Nik,NikRep,NikRR,China,ChinaCom,Sok,SokCom,SokComRep,Berge,BergeCom,BergeRep,Wiesn,Yuan1} after the first experiment with photons \cite{Danan} confirmed discontinuous trajectories.

\begin{figure}
\begin{center}
  \includegraphics[width=8cm]{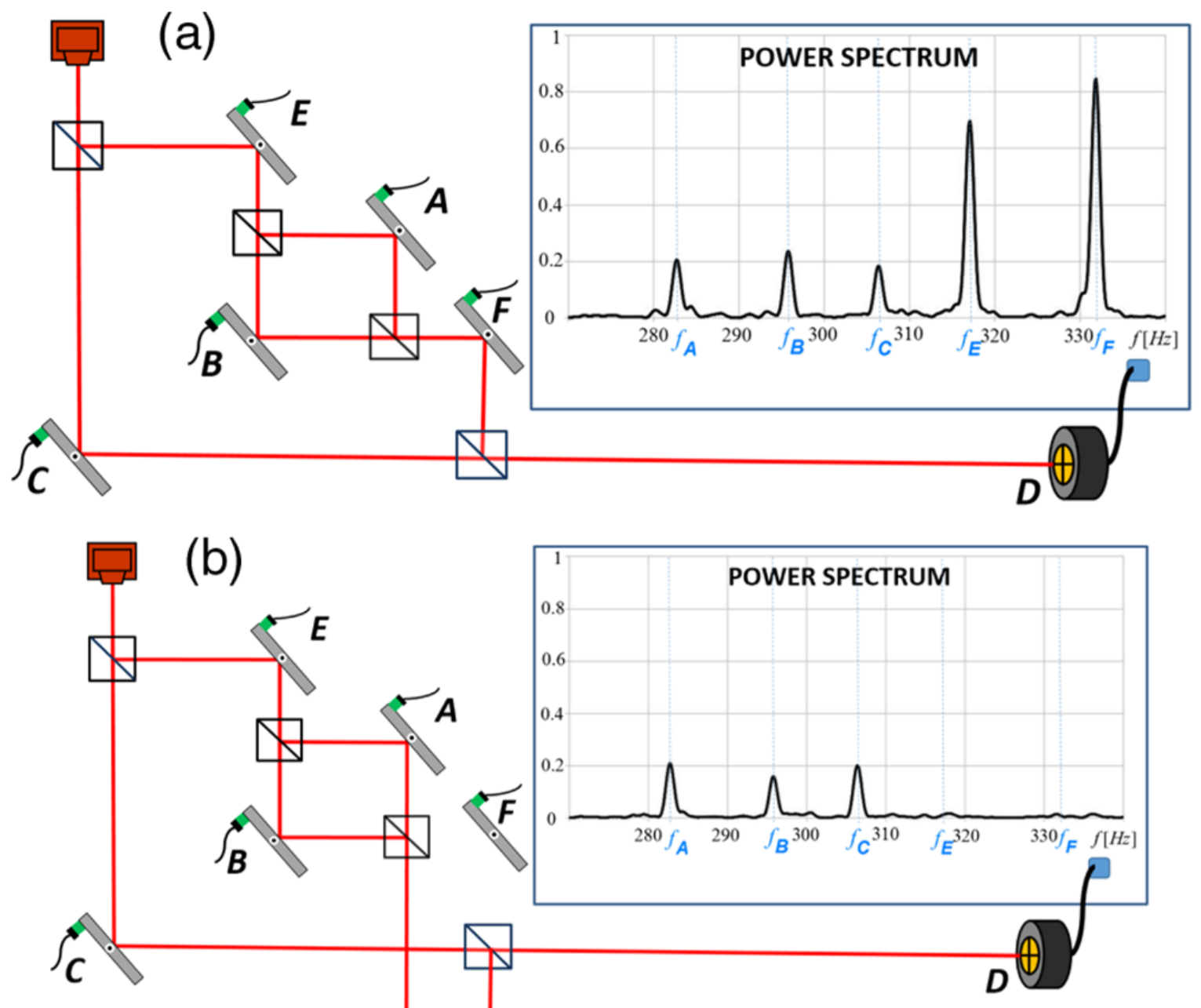}
\end{center}
  \caption{
 Measuring faint traces in an optical nested Mach-Zehnder interferometer. a) Constructive interference. b) Destructive interference in the inner interferometer towards $F$. (Figures 3a and 3b in  \cite{Danan}).
  }
  \label{fig:FIG-1}
\end{figure}
\begin{figure}
\begin{center}
  \includegraphics[width=8cm]{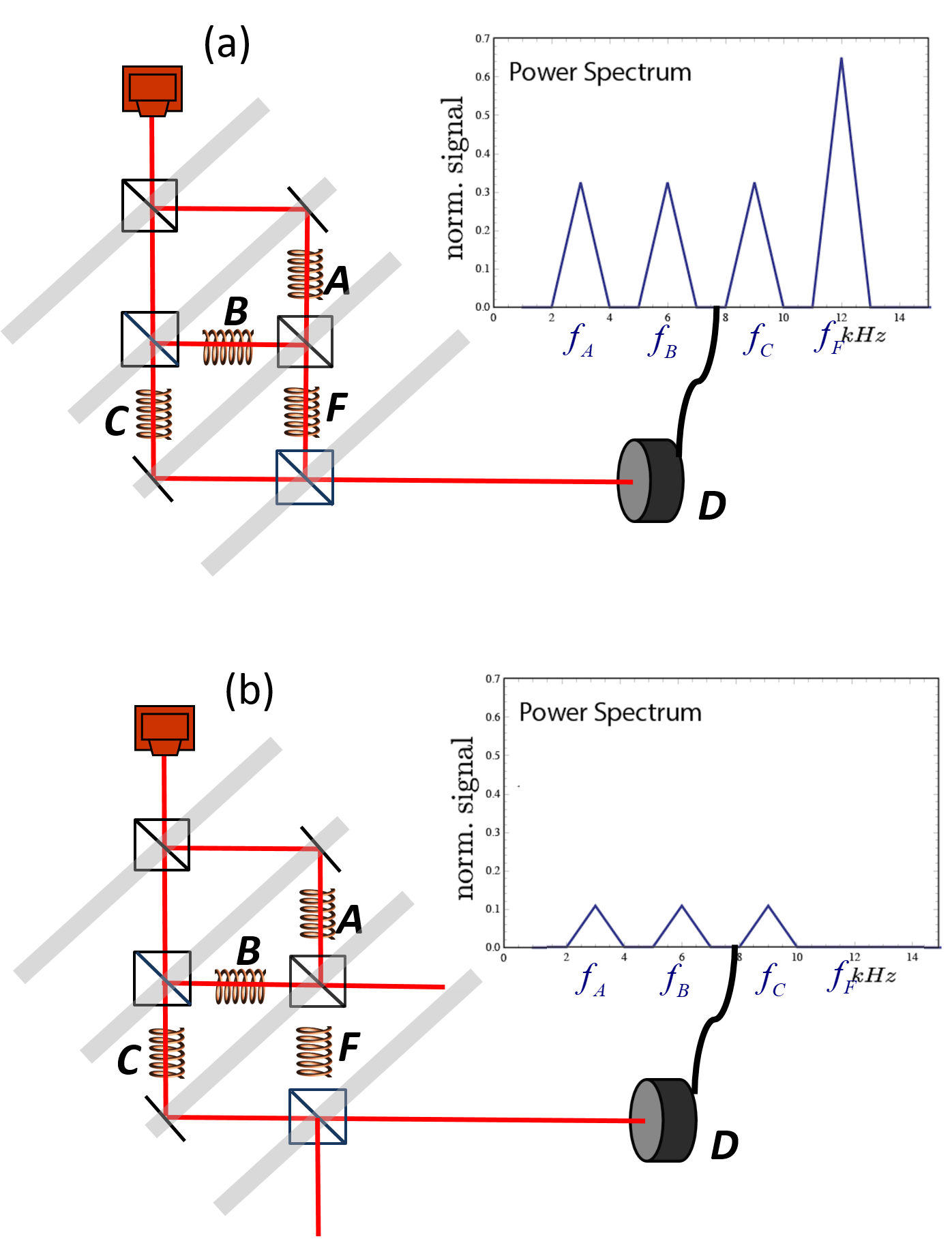}
\end{center}
  \caption{(a) A schematic description of the experimental setup of the  neutron interference experiment with simulation of the results (taken from  \cite{AskNeut},  Fig. 3). (a) Constructive interference. (b) Destructive interference in the inner interferometer towards $F$.
  }
  \label{fig:FIG-2}
\end{figure}

In the optical experiment \cite{Danan}, the nested MZI was tuned in two particular ways, see Fig.~1. In the first run, performed as a test of the system, there was constructive interference  in both the inner and the external interferometers. The faint traces were observed in all segments of the interferometer, $A$, $B$, $C$, $E$, and $F$, see Fig.~1 (a). In the second run, the inner interferometer was tuned to create destructive interference toward the final beam splitter of the external interferometer and the faint traces were observed on path $C$, but also inside the inner interferometer on paths $A$ and $B$. No traces were observed on the paths leading towards and out of the inner interferometer, segments $E$ and $F$, see Fig.~1 (b).

\section{Neutron interference experiment}
An experiment with neutrons \cite{Hasegawa} recently reported in a paper \cite{AskNeut}  ``Asking neutrons where they have been'' (following the title of \cite{Danan} ``Asking photons where they have been'') ends with the conclusion: ``The discontinuous trajectories
of the photons described in  \cite{Danan} stem from a misinterpretation of the observed faint traces. A
faint trace is only a sufficient but not a necessary condition for the presence of the particle.
Our experimental results witness the multifold presence of the neutron's wave function in the
interferometer.'' The authors, do not state explicitly where was the neutron in their experiment. However, from the phrase ``A
faint trace is only a sufficient but not a necessary condition for the presence of the particle'' it is clear that they admit the presence of the neutron inside the inner interferometer (contested by others, e.g. \cite{Berge}). Then, to avoid ``discontinuous trajectories'' the neutron has to be in all arms of the interferometer.

A small difference  in the neutron interference experiment, Fig. 2. was that it had no segment $E$. To make an exact copy of the optical experiment one would require a crystal with five plates, instead of four which were used. (Even four plates is a technological achievement, since most previous neutron interference experiments had just three plates.) A bigger difference was in the experimental method of marking the paths. In the optical experiment \cite{Danan}, the markings were given by frequencies of the vibrations of the mirrors in the interferometer. Rotation of the mirror provided a tiny change in the transverse momentum of the photon which correspondingly influenced the transverse position of the detected photons. In the neutron experiment, by contrast, the marking was done by placing  resonance-frequency spin-rotators  operating at different frequencies in various arms of the interferometer. Neutrons passing through different arms obtained different  energies  which  were detected by analyzing energy spectrum of the neutrons.

\section{Optical counterpart of neutron interference experiment}
Another optical experiment  \cite{Zhou}  used a  which-path marking method, conceptually identical to that of the neutron interferometer, see Fig. 3,  so it is possible to perform a direct comparison between optical and neutron interference experiments. Various electro-optic phase modulators (EOM) were introduced in different paths of the original nested MZI. They caused various frequency sidebands which were detected in spectrum analyses of the detected photons.  The authors  of this optical experiment reported full agreement with the conclusions of \cite{Danan}. They observed ``anomalous trajectories which are noncontinuous and in some cases do not even connect the source of the photon to where it is detected''.

In contrast, the authors of the neutron interference experiment \cite{Hasegawa,AskNeut} criticised the conclusions of \cite{Danan} and \cite{Zhou} in spite of the fact that their reported results actually support the surprising discontinuous picture of the past of the particle in the nested MZI. The neutron experiment had significant systematic errors which made it difficult to interpret their experimental results, but the authors presented simulations of expected results for an ideal case, see Fig. 2. The simulation of the power spectrum from their neutron interference experiment exactly fits (except for missing arm $E$) the actual data of the optical experiment \cite{Zhou}, see Fig. 3.
\begin{figure}
\begin{center}
  \includegraphics[width=8cm]{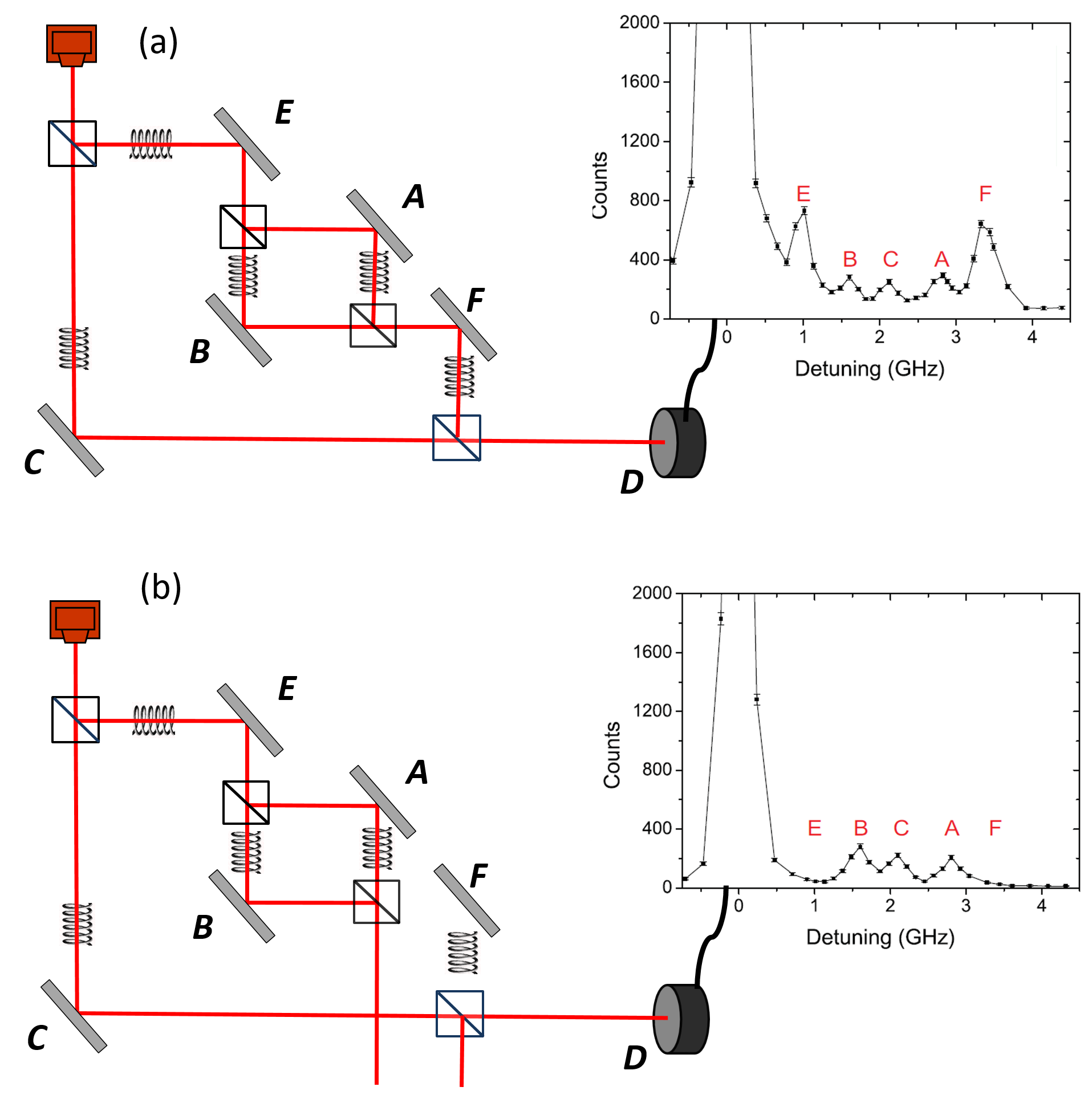}
\end{center}
  \caption{Optical nested Mach-Zehnder interferometer
with electro-optical modulators having different modulation frequencies, placed in every arm.
(a) Constructive interference. (b) Destructive interference in the inner interferometer towards $F$.
  }
  \label{fig:FIG-3}
\end{figure}
Both in \cite{Hasegawa} (as I demonstrated in an unpublished comment \cite{myCom}) and in the recent report \cite{AskNeut}, the authors of the neutron interference experiment clearly state that the ``faint traces'' they observed are the which-path markers of the neutron. The results, or at least their simulation of the results, were presented in their Fig. 3 (repeated here in Fig. 2.) as a ``which-way signal''. The graphs   clearly show disconnected trajectories  predicted in \cite{past} and observed in \cite{Danan} and \cite{Zhou}.

Let us discuss in more detail the methods of measuring the faint trace. In all experiments \cite{Danan,Zhou,AskNeut} the measuring device was the particle itself. The faint trace record was ``written'' on the degree of freedom of the particle and it had to remain intact until the particle reached the detector which ``read''   the record.  In experiments \cite{Danan} the faint trace record was written on the transverse momentum and it was important to pay special attention so as not to distort it. For example, adding a dove prism to the inner interferometer makes the experiment incorrect \cite{Jordan,JordanCom}.
In the  optical experiment \cite{Zhou} and in the neutron experiment \cite{Hasegawa} the information about the faint trace was written in the energy of the particle. This kind of record is more robust; a dove prism would not spoil the experiment.

 On one hand, the optical experiment \cite{Zhou}, which is a single-particle experiment  seems to be better, more ``quantum'' than the original  experiment \cite{Danan} performed with laser light. On the other hand, the single-particle experiment  has a conceptual difficulty. Every detection of the modified energy of the particle  corresponds to a strong measurement which provides unambiguous path determination. All such particles had known continuous trajectories. The task, however,  is to determine where were the particles inside the interferometer without strong which-path measurement. This was deduced through indirect inference from many detection events in the ensemble. The statistics of  detection events  at different sideband energies provided information for the majority of photons which were not disturbed and were detected with the basis energy.  It was assumed that these undisturbed photons were (sometimes simultaneously) in arms where a significant (although still very small) fraction of photons were disturbed and then detected, and that there were no photons (disturbed or undisturbed) in arms where no disturbed  photons were found. In contrast, in the experiment with vibrating mirrors no information was obtained about individual particles, so there were no photons with a known continuous path.

\section{Which-path measurement in Mach-Zehnder interferometer}

It is claimed that another which-path experiment in MZI  \cite{Berge2} ``has an immediate bearing on the experiments by
Danan et al. \cite{Danan} and Zhou et al. \cite{Zhou}.''  In this experiment, the part of the original interferometer, the inner tuned interferometer, was analyzed. It was shown that when a particle was detected at a dark port of the MZI, then robust which-path information was present in the environment.  In their words \cite{Berge2}:``every photon that emerges from the dark output port of the balanced interferometer has a known path''. In the experiment they demonstrated a different result. They showed that photons with a known path end up in a dark port and that {\it undisturbed} photons with unknown path did not reach the dark port.
The sentence: ``every photon that emerges from the dark output port of the balanced interferometer has a known path'' has to be clarified. It  seems that in general it is not correct, and even in the most favorable conditions its meaning is limited. In an ideal balanced interferometer there are no photons reaching the dark port. In the context of our discussion, I assume that ``balanced interferometer'' is different from an ideal balanced interferometer due to decoherence, i.e. that the photon leaves a trace in the environment, but the state of the photon moving  in any one arm is not affected. Let us consider the case of an equal decoherence strength in all arms. In this case the  quantum states of the local environment in the arms, $|\Phi_A\rangle$, $|\Phi_B\rangle$ are not changed when the photon is not present and the presence of the photon changes them in the following way:
\begin{equation}\label{1}
  |A\rangle|\Phi\rangle_A \rightarrow |A\rangle|(\eta|\Phi_A\rangle+\epsilon|\Phi_{A\perp}\rangle),
\end{equation}
\begin{equation}\label{2}
  |B\rangle|\Phi\rangle_B\rightarrow |B\rangle (\eta|\Phi_B\rangle + \epsilon|\Phi_{B\perp}\rangle).
\end{equation}
In this scenario, for the case of the detection of the photon in the dark port, the final state of the environment will be
\begin{equation}\label{3}
 \frac{1}{\sqrt 2}(|\Phi_{A\perp}\rangle|\Phi_B\rangle-|\Phi_A\rangle|\Phi_{B\perp}\rangle).
\end{equation}
This is what apparently corresponds to a ``known path'' - the  quantum state of the environment is in complete correlation with the path (this would not be the case if  the couplings were different in the two arms).

However, complete correlation does not correspond to the known path in the usual sense. ``Known'' path is a single definite path, which some observers might not know, but for which there are unambiguous  records somewhere. In our situation, even if we are given complete information about the system and the environment,  we do not know if the path was $A$ or $B$, in some sense it was both.

The states of the environment (\ref{3}) might well remain at the microscopic quantum level for a long time. Then we can consider a quantum erasure scenario \cite{erasure} in which a measurement of the environment, say in arm $A$, is performed immediately after a photon interacts with the environment,  in the basis $|\pm\rangle$:
\begin{equation}\label{4}
 |\pm_A\rangle\equiv\frac{1}{\sqrt 2}(|\Phi_{A\perp}\rangle \pm|\Phi_A\rangle ).
\end{equation}
Now, at the time the photon is detected in the dark port, we already know that the state of the environment is either $|+_A\rangle |-_B\rangle$ or $|-_A\rangle |+_B\rangle$. Both cases correspond  to the situation in which we have no information about the path the photon took. Note that  our quantum erasure measurement does not change the motion of the photon if it is well localized in one of the arms, so we can still claim that this is an experiment with a balanced interferometer. Thus, my  proposal provides a counterexample to the statement ``every photon that emerges from the dark output port of the balanced interferometer has a known path''.

The main reason why experiment \cite{Berge2} does {\it not} ``have an immediate bearing on the experiments by Danan et al. \cite{Danan} and Zhou et al. \cite{Zhou}'' is that the photons in the latter experiments were  not detected in the dark port. We can only infer the presence of the photon in the dark port using  classical logic, the legitimacy of which (in its application to quantum particles) was challenged in  the work on the past of the photon \cite{past}. The questionable classical argument is:  photons which were in the inner interferometer were also  in its  dark port, because otherwise they could not reach the final detector of the nested interferometer. The proof that the trajectory is continuous is not legitimate if it presupposes that the trajectory must be continuous \cite{BergeCom}.

\begin{figure}
\begin{center}
  \includegraphics[width=7cm]{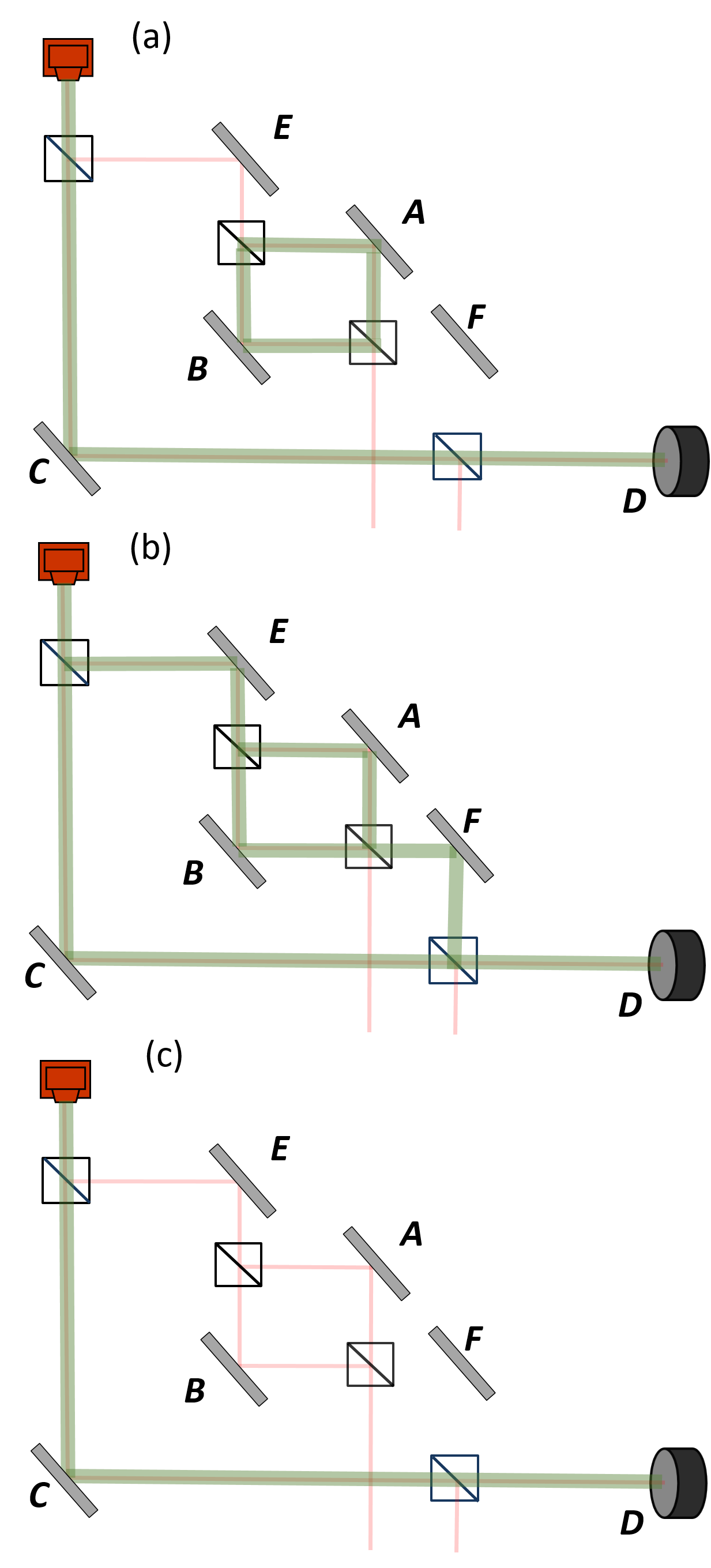}
\end{center}
  \caption{The controversy about where were the particles (thick green line)  in the tuned nested Mach-Zehnder interferometer. (The red line shows the wave function of the particle.) (a) the discontinuous trajectory proposed in \cite{past} and confirmed in \cite{Danan,Zhou} (b) The continuous trajectory in all arms of the interferometer, claimed to be confirmed  by a neutron interference experiment  \cite{Hasegawa,AskNeut}. (c) A single continuous trajectory, the ``common sense'' picture advocated by \cite{Berge} and
  claimed to be supported by the experimental results \cite{Berge2}.
    }
  \label{fig:FIG-4}
\end{figure}

\section{The controversy about nested Mach-Zehnder interferometer}

The controversy about where the particles have been in the nested interferometer is the controversy of the three competing pictures presented in Fig. 4: (a) Discontinuous trajectories with a presence in $A$, $B$, and $C$, but not in $E$ and $F$, proposed in \cite{past} and confirmed in experiments as the location of the first order faint traces left by the pre and postselected particle.  (b) Everywhere inside the interferometer $A$,$B$, $C$,  $E$ and $F$. This picture is advocated by the authors of the neutron interference experiment. (c) A continuous single trajectory through arm $C$ corresponding to classical common sense as the only possible path of the wave packet from the source to the detector. Is there a way to decide which picture correctly describes where the pre and postselected particles have been? No! There is no consensus about the independent definition of where the particle was, one can only argue which definition is more natural and more useful.  Note, that in Bohmian mechanics \cite{Bohm} the trajectory is different from all three pictures of Fig. 4, see Fig. 5  \cite{surr}.

 In my view picture (c), although  very natural according to classical common sense - the only path the particle could take - is deeply misleading. The single trajectory passing through $C$ would be confirmed if an unambiguous path discrimination measurement that path $C$ is the trajectory were to be performed. But it is definitely wrong to consider it as the only trajectory of the pre and postselected particle without unambiguous  path discrimination measurements, since a continuous path passing through $A$ would also be confirmed with certainty if an unambiguous test were to be performed there instead.

\begin{figure}
\begin{center}
  \includegraphics[width=7cm]{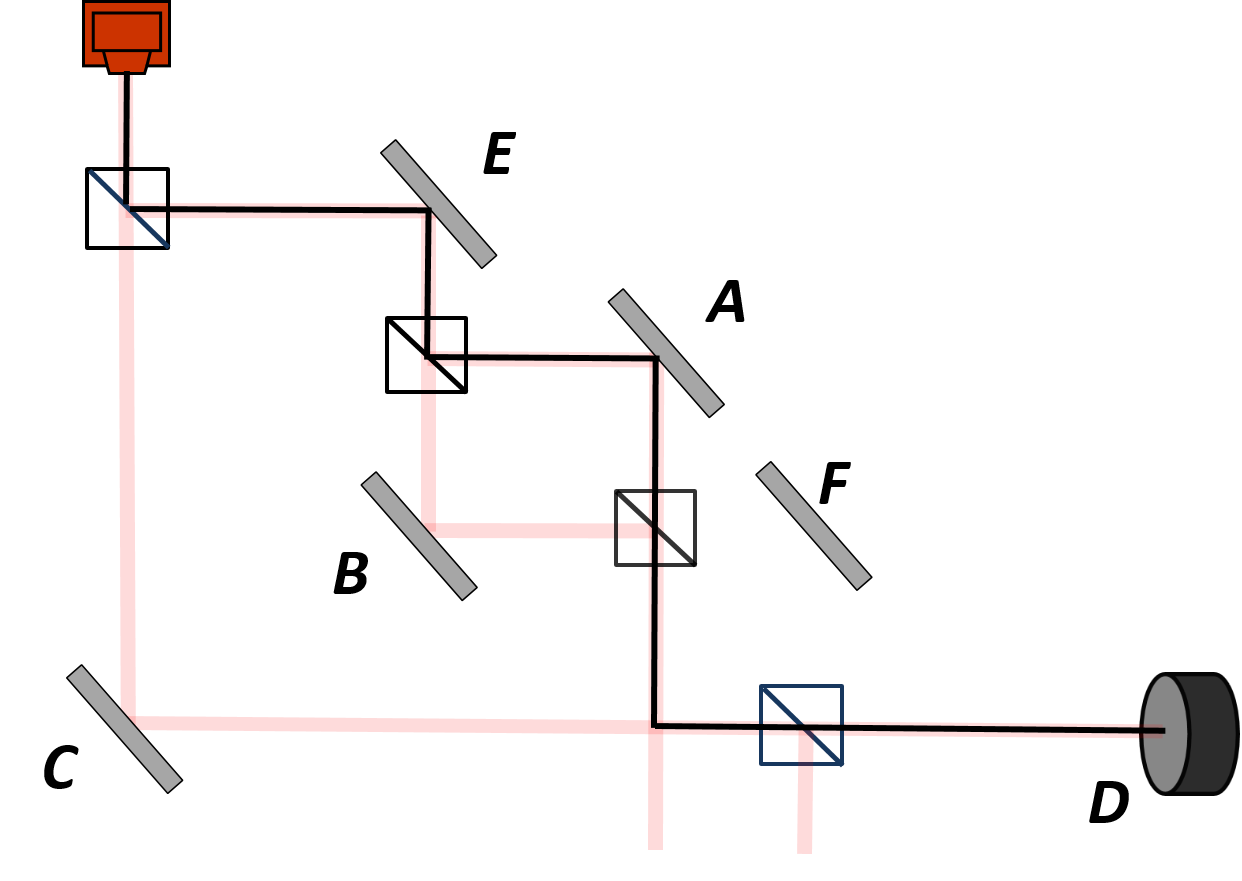}
\end{center}
  \caption{Bohmian trajectory (black  line) of the particle detected by detector $D$ in the tuned nested Mach-Zehnder interferometer. (The red thick line shows the wave function of the particle.)  }
  \label{fig:FIG-5}
\end{figure}

Picture (b) also seems to be misleading and not useful for our purposes. It is not clear  why the authors of the neutron interference experiment advocated this picture, but one way for its justification is an observation that presence of the trace inside the inner interferometer in arms $A$ and $B$ cannot happen without the presence of some trace in $E$ and in $F$. However, this picture does not distinguish between areas with faint traces, similarly to the traces of a localized particle, and areas with much smaller faint traces. The ratio between the traces in $A$, $B$, and $C$  and the traces in $E$ and $F$ might be arbitrarily large. (Note the concept of a {\it secondary} presence I introduced in \cite{morepast}.)  In particular, this picture seems to be impractical as it implies the nonexistence of dark ports in interferometers.

 I find  the faint trace picture (a) the most natural and useful. What helps me is the two-state vector formalism (TSVF) \cite{TSVF} which describes pre and postselected particles by the two-state vector consisting of the usual quantum state  evolving forward in time and the other one, evolving backward in time from the detector. The TSVF provides a  very simple proof for a natural  counterfactual property of picture (a). In every location the particle is not present according to the trace criterion (a), there is zero probability to find it in a strong measurement (provided it is only searched for there). Indeed, if in a particular place there is no first order weak trace, the weak value of the projection on that place must be zero. The projection operator is a dichotomic variable, and for such variables, according to the theorem introduced in \cite{AV91}, if the weak value equals an eigenvalue, this eigenvalue is obtained in a strong measurement with certainty. Therefore, we know with certainty that in arms $E$ and $F$ directed towards and out of the inner interferometer, we will not find the particle in a strong nondemolition measurement.

 Another nice feature of the TSVF is that it admits a formal definition of the weak trace  picture  (a): the overlap of the forward and backward evolving states. The discontinuity of the particle trajectory then gets a simple explanation. Both forward and backward evolving states exhibit continuous time evolution, but the overlap might exhibit discontinuities.

 Finally, the TSVF can upgrade the picture (a) of the particle's path to the picture which describes not only {\it where} but also {\it how} the particle was present.  In general, the weak value of the projection can be any complex number. In \cite{past} $({\bf  \rm P}(\vec{r}))_w=0$ corresponds to a particle not being present in $\vec{r}$ while $({\bf  \rm P}(\vec{r}))_w\neq0$ (and it is not a very small number) corresponds to a particle  being present in $\vec{r}$. The particle either affects the environment in $\vec{r}$ or it does not affect it. A deeper understanding of the concept of the weak value \cite{weakPRA17} tells us that the weak value of the projection operator describes the universal modification of all local weak interactions \cite{PNAS}. All couplings might be amplified, reduced, changed in the direction of their action, and, if the weak value is imaginary, changed in the type of actions that are effected: a shift instead of a kick, a kick instead of a shift. Thus,  $({\bf  \rm P}(\vec{r}))_w$ tells  us where and how the particle is present. In our case $({\bf  \rm P}_C)_w=({\bf  \rm P}_A)_w=1$, $({\bf  \rm P}_E)_w=({\bf  \rm P}_F)_w=0$, and $({\bf  \rm P}_B)_w=-1$. The experiments did not observe the opposite direction of the disturbance in arm $B$, because they were sensitive only to the size of the signal.

 Precise details of small interactions with the environment in a tuned nested interferometer will allow for calculation of the tiny weak values of projections on $E$ and $F$, providing a precise picture of the pre and postselected particle in all arms of the interferometer, including the differences between a full and secondary presence. Clearly, it will be superior to the less informative picture (b).

\section{Conclusions}

In the words of one of the referees, the answer to the question in the title of this paper is: ``Of course
not! After all, interferometers, whether optical or neutron, work
very much on the same principles and both particles obey the same
quantum mechanical laws, so there should not be any discrepancies.''
There have been, however,  controversies in the published literature
concerning this matter and the present paper attempts to clarify the
situation.

I described four experiments. The original experiment with a particularly tuned optical nested MZI demonstrating a peculiar pattern of faint traces the photons leave. Another optical experiment considered the same configuration, but was performed with single photons and a different method of observing the traces. I also considered a neutron experiment involving a nested MZI, analogous to the single photon experiment, and finally a single photon experiment of a single MZI tuned to exhibit complete destructive interference in one of the ports as the inner MZI of the original experiment. I have shown that indeed there are no discrepancies between the experimental data collected in all these experiments.

The data confirms standard quantum mechanics, but it also supports a particular interpretation of these experiments according to which the location of the pre and postselected particles in the setups of nested interferometers is not described by continuous trajectories starting at the source and ending at the detector. This interpretation is based on the definition: the particle was where it left a faint trace. This definition is not part of standard quantum mechanics, which is silent regarding the question of which path was taken by a pre and postselected quantum particle.

In all the experiments I discussed, the measuring device testing where the particle left a trace  inside an interferometer  was the particle itself. A conceptually better experiment would be a direct observation of the traces left by the particles using an external measuring device \cite{ChinaCom}. Such an experiment is more challenging, but seems to be feasible in a foreseen future (see, for example, the case of a weak measurement with an external measuring device \cite{Steinberg}).

The current analysis sheds light on the question of the validity of the criticism of the paper \cite{past}. In this paper the presence of particles is {\it defined} according to the first order trace (i.e. similar to the trace a well-localized packet leaves), and this trace can be analyzed theoretically and experimentally. I argued there that this criticism is incorrect, and that all experiments testing faint traces including  \cite{AskNeut} actually support the discontinuous picture \cite{past}, while the experiment of Ref. \cite{Berge2} is not relevant. This is firstly because it is not an experiment with a nested interferometer, and secondly because it is a strong measurement with unambiguous path discrimination. My analysis also questions the validity of the main conclusion of \cite{Berge2}: ``every photon that emerges from the dark output port of the balanced interferometer has a known path''.

The validity of standard quantum mechanics is not brought into question by my analysis and the experimental data of all discussed experiments is fully consistent with it.  However, I find that the continuous effort to extend the range of concepts of quantum theory (here: asking the question of where a pre and postselected particle has been) is fruitful. It brings with it new intuitions and a deeper understanding of nature which enables the progress of quantum technology.

I thank Stephan Sponar for helpful correspondence. This work has been supported in part by the Israel Science Foundation Grant No. 2064/19 and the National Science Foundation - U.S.-Israel Binational Science Foundation
Grant No. 735/18.







\end{document}